\documentclass[twocolumn,aps,prl,superscriptaddress,nofootinbib]{revtex4}

\usepackage{amssymb,xcolor}
\usepackage{epsfig,amssymb,amsmath}

\newcommand{\ket}[1]{|#1\rangle}

\newcommand{\eq}{\begin{equation}}
\newcommand{\fine}{\end{equation}}

\newcommand{\aac}{\'a}

\newcommand{\fisicarm}{Dipartimento di Fisica, Sapienza Universit\'a di Roma, I-00185 Roma, Italy}
\newcommand{\ino}{Istituto Nazionale di Ottica, Consiglio Nazionale delle Ricerche (INO-CNR), L.go E. Fermi 6, I-50125 Firenze, Italy}
\newcommand{\cfermi}{Museo Storico della Fisica e Centro Studi e Ricerche Enrico Fermi, Via Panisperna 89/A, Compendio del Viminale, I-00184 Roma, Italy}
\newcommand{\ifn}{Istituto di Fotonica e Nanotecnologie, Consiglio Nazionale delle Ricerche (IFN-CNR),
Piazza L. da Vinci, 32, I-20133 Milano, Italy}
\newcommand{\fisicami}{Dipartimento di Fisica, Politecnico di Milano, Piazza L. da Vinci, 32, I-20133 Milano, Italy}

\begin{document}

\title{Polarization entangled state measurement on a chip}

\author{Linda Sansoni}
\affiliation{\fisicarm}
\author{Fabio Sciarrino}
\affiliation{\fisicarm}
\affiliation{\ino}
\author{Giuseppe Vallone}
\affiliation{\cfermi}
\affiliation{\fisicarm}
\author{Paolo Mataloni}
\affiliation{\fisicarm}
\affiliation{\ino}
\author{Andrea Crespi}
\affiliation{\ifn}
\affiliation{\fisicami}
\author{Roberta Ramponi}
\affiliation{\ifn}
\affiliation{\fisicami}
\author{Roberto Osellame}
\affiliation{\ifn}
\affiliation{\fisicami}

\begin{abstract}
The emerging strategy to overcome the limitations of bulk quantum optics consists of
taking advantage of the robustness and compactness
achievable by the integrated waveguide technology. Here we report
the realization of a directional coupler, fabricated by femtosecond laser waveguide
writing, acting as an integrated beam splitter able to support polarization encoded
qubits. This maskless and single step technique allows to realize circular transverse waveguide
profiles able to support the propagation of Gaussian
modes with any polarization state.
Using this device, we demonstrate the quantum interference
with polarization entangled states and singlet state projection.
\end{abstract}

\maketitle

Photons are the natural candidate for Quantum Information (QI)
transmission \cite{gisi02rmp, duan01nat}, quantum computing \cite{ladd10nat, kok07rmp}
optical quantum sensing and metrology \cite{dowl08cph}.
However, the current optical technology does not allow the transition
to ultimate applications for many practical limitations.
Complex quantum optical schemes, realized in bulk optics
suffers from severe limitations, as far as stability, precision and
physical size are concerned. Indeed, it is a difficult task to build
advanced interferometric structure using bulk-optical
components with the stability and the optical phase control accuracy
necessary to reach the sensitivity allowed by quantum mechanics.
Furthermore, it is very difficult to reach this task outside
environments with controlled temperature and vibrations, and this
{makes} applications outside laboratory hard to achieve.
\begin{figure}[t]
\centering
\includegraphics[width=6cm]{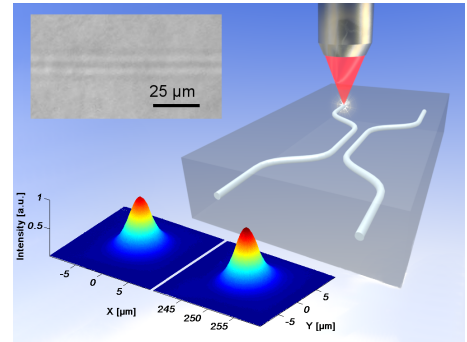}
\caption{Schematic of the femtosecond-laser-written
directional coupler in the bulk of a borosilicate glass. Upper inset
shows a microscope image of the two waveguides in the coupling
region. Lower inset shows the near-field intensity profile of the
output guided modes of the directional coupler by launching light in
a single input; the symmetric Gaussian shape and the balanced
splitting in the two arms can be appreciated.}
 \label{setup}
\label{fig:ulwbs}
\end{figure}

The present approach to beat these limitations is to adopt
miniaturized optical waveguide devices. Very
recently it was reported \cite{poli08sci,poli09sci} that
silica waveguide circuits integrated onto
silicon chips can be successfully used to realize key
components of quantum photonic devices. Inherently stable
interferometers were shown to demonstrate phase stability, not only
of single path encoded qubits, but also of two-photon entangled Fock
state. On this basis, miniaturized integrated quantum circuits were
realized to implement the first integrated linear optical control-NOT
gate, achieving a fidelity very close to the theoretical value
\cite{poli08sci}. More recently, novel components for adaptive
quantum circuits have also been demonstrated \cite{matt09npho}.
These experiments demonstrate
robust and accurate phase control in integrated,
path-encoded waveguide systems. Similar results have been
obtained in UV laser written optical circuits fabricated in a suitable
stack of doped silica layers on a silicon substrate \cite{smit09ope}.

All the experiments performed so far with integrated quantum circuits
are based only on path encoded qubits with a given polarization state of the photons.
On the other hand, many QI
processes and sources of entangled photon states are based on the
polarization degree of freedom \cite{kok07rmp}. One important example
is given by states built on many photons \cite{kris10npho} and/or
many qubits, and by several schemes of one-way optical
quantum computing \cite{walt05nat}. Hence it is of essential interest
to include the use of photon polarization in quantum
circuits by fabricating integrated polarization independent devices,
i.e. able to efficiently guide and manipulate photons in any
polarization state.

It has to be noticed that the above mentioned
silica-on-silicon and UV written integrated waveguides suffer from
intrinsic birefringence (usually reported in the order of $4 \times
10^{-4}$ \cite{kawachi90oqe,johlen00jlt}). In fact, these waveguides
are fabricated in a doped silica multilayer structure on a silicon
substrate and this causes material stress due to lattice mismatches
between the different layers. Techniques for reducing this stress and
the induced birefringence have been proposed, but they pose serious
difficulties in terms of fabrication complexity and reproducibility
\cite{inoue97jlt}. Such birefringence causes polarization-mode
dispersion and results in polarization dependent behavior of the
integrated devices, which removes indistinguishability between the
two polarizations. Moreover, propagation in birefringent structures
can cause decoherence of large-bandwidth (short coherence time)
photons typically generated in parametric down-conversion
experiments. As a
consequence, the techniques already employed for producing
path-encoded quantum circuits are not appropriate for processing
polarization-encoded qubits in integrated devices.

In the present paper we show how to guide and manipulate photons in
any polarization state by adopting a recently introduced technique,
based on the use of ultrashort laser pulses, for direct writing of
photonic structures in a bulk glass
\cite{gattass08np,dellavalle09joa}. Precisely, here,
for the first time, we demonstrate the maintenance of polarization
entanglement and Bell-state analysis
in an integrated symmetric (50/50) beam splitter, opening the way to
the use of polarization entanglement in integrated circuits for QI
processes.

\begin{figure}[t]
\centering
\includegraphics[width=9cm]{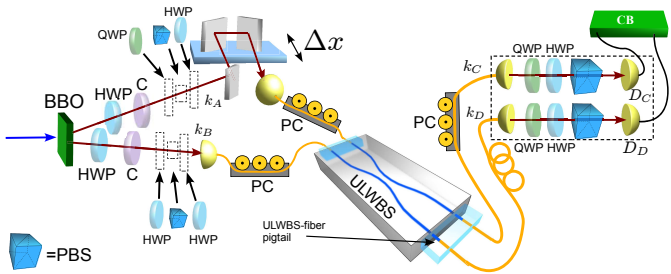}
\caption{
Setup for the quantum optics experiments showing the source of polarization-entangled photons, the ULWBS and the detection system. Polarizing beam splitter (PBS), half (HWP) and quarter (QWP) waveplates were optionally inserted in path $k_A$ and $k_B$ to prepare different input states.
A delay line $\Delta x$ in the $k_A$ arm enabled temporal delay variation between the two input photons.
The components shown in the dashed box were inserted only during the tomography measurement of the filtered state.
C: crystal compensators, PC: polarization controllers.
}
\label{fig:setup}
\end{figure}
Direct fabrication of buried waveguides in glass is obtained by
femtosecond laser micromachining. Femtosecond infrared pulses,
focused into the substrate using a microscope objective, induce
non-linear absorption phenomena based on multiphoton and avalanche
ionization. These processes lead to plasma formation and energy
absorption in a small region confined around the focus, causing a
permanent and localized modification of the bulk material. Adjusting
the processing parameters a smooth refractive index increase can be
obtained and light-guiding structures are produced by translating the
substrate with respect to the laser beam. Microscopic mechanisms
leading to refractive index increase are complex and include
densification, structural modification, color centers formation,
thermal diffusion and accumulation. They concur in different ways
depending on the specific material and fabrication parameter
combination, i.e. wavelength, duration and energy of the laser
pulses, repetition rate, objective numerical aperture (NA) and
translation speed \cite{osellame06jstqe}.

The ultrafast laser writing approach has several
advantages: i) it is a maskless technique, thus particularly suited
for rapid prototyping of devices; ii) it can easily fabricate buried
optical waveguides in a single step; iii) it can produce optical
circuits with three-dimensional layouts; iv) it can provide
waveguides with a circular transverse profile \cite{osel03jos}, that
can support the propagation of Gaussian modes with any polarization
state, with very low waveguide form birefringence.
Ultrafast laser written (ULW) waveguides in fused
silica substrate have been recently employed for quantum optics
experiments, still with path-encoded qubits \cite{mars09ope}.
However, it is known that ULW waveguides in fused silica are affected
by material birefringence \cite{cheng09ope} (in particular when high
refractive index changes are required, as in the case of curved
waveguides) due to the formation of self-aligned nanogratings in the
material during the irradiation process \cite{shimotsuma03prl}.
Moreover fabricating waveguides in fused silica is a rather slow
process (in the order of $10-100 \mu$m/s) \cite{ams08jstqe}. For
these reasons we chose to employ a borosilicate glass (EAGLE2000,
Corning) as substrate, where the formation of nanogratings has never
been observed \cite{ams08jstqe}. In addition, high repetition rate
laser pulses induce
isotropic thermal diffusion and melting of the material around the
focal point \cite{eaton08ope}, providing almost circular waveguide
cross-section without the need for any shaping of the writing laser
beam. Very low-loss waveguides are obtained with translation speeds
as high as 1-5 cm/sec, allowing extremely short processing times \cite{si1}.
This represents an advantage for the realization of complex photonic
circuits.

At wavelengths around $800 $nm the waveguides support a single
Gaussian mode of circular profile with $8 \mu$m diameter at $1 / e^2$
({see the near-field intensity profile of the guided modes in the
lower inset of Fig.} \ref{fig:ulwbs}), allowing a $85\%$ overlap
integral with the measured mode of the fiber used (Thorlabs
SM800-5.6-125) and leading to 0.7 dB estimated coupling losses.
Measured propagation losses are $0.5 $dB/cm and using a curvature
radius of $30 $mm additional bending losses are lower than $0.3
$dB/cm. The birefringence of the ULW waveguides has
also been characterized \cite{si1}, providing a value $B = 7
\times 10^{-5}$, thus about one order of magnitude lower than
silica-on-silicon waveguides.

Ultrafast laser written beam splitters (ULWBS) were fabricated with
the {directional coupler} geometry, as shown in Fig. \ref{fig:ulwbs}.
Straight segments and circular arcs of $30 $mm radius were employed
for an overall device length of $24$ mm. Waveguides
starts with a relative distance of $250\mu$m and in the interaction
region they get as close as $7\mu$m (see upper inset in Fig.
\ref{fig:ulwbs}). This distance is the smallest one avoiding overlap
between the two waveguides. This choice was made to minimize the
sensitivity to fabrication imperfections and to obtain the shortest
possible interaction length, given that future quantum optic devices
will require several cascaded components integrated in the same chip.
In order to optimize the length $L$ of the central straight segments
several directional couplers have been fabricated varying such length
($L= 0\div1000 \mu$m) and the corresponding splitting ratios were
measured. $L = 0 \mu$m is the shortest length yielding a
splitting ratio of about $50\%$ (see the ULWBS output modes in the
lower inset in Fig. \ref{fig:ulwbs}) at $806$nm wavelength. Indeed,
the possibility of achieving a $50\%$ splitting with no straight
segments is due to the coupling between the modes already occurring
in the curved parts of the two approaching/departing waveguides.
\begin{figure}
\centering
\includegraphics[width=8.3cm]{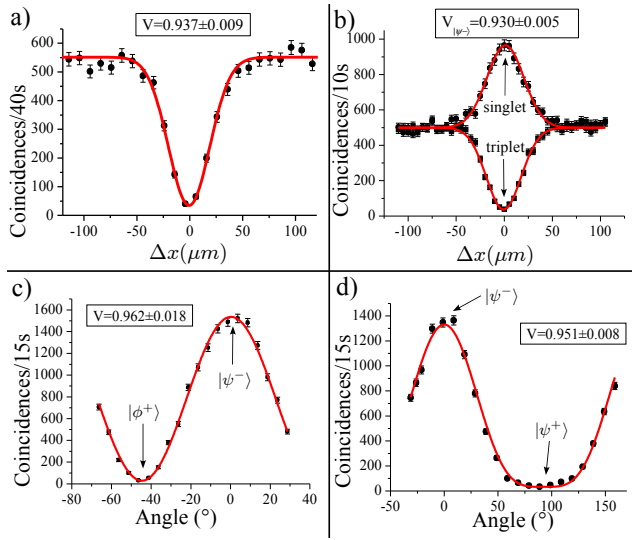}
\caption{a) HOM dip with input state $\ket{HH}$. b) peak/dip corresponding to the singlet/triplet input state.
c) Fringe pattern obtained by rotating the HWP on mode $k_B$.
d) Fringe pattern obtained by rotating the QWP on mode $k_A$.
All curves represent experimental fits.}
\label{fig:osc}
\end{figure}The reflectivity of the ULWBS for the horizontal and vertical
polarizations was measured with a tunable laser operating at $806$nm.
The measured unbalancement between the two reflectivities
$R_H=(49.2\pm0.2)\% $ and $R_V= (58.1\pm 0.2)\%$, is
attributed to a residual ellipticity in the waveguide
profile, notwithstanding the thermal mechanism of the waveguide
formation. Work is in progress to further optimize the waveguide
cross-section with the astigmatic beam shaping technique
\cite{osel03jos}.

We demonstrated the ability of the chip to preserve any incoming
polarization state by measuring the polarization degree (G) and obtaining $G\geq99.8\%$.

The suitability of the ULWBS to handle polarization encoded qubits
was demonstrated by manipulating polarization
entangled states. The four Bell states
$\ket{\psi^\pm}=\frac1{\sqrt2}(\ket H_A\ket{V}_B\pm\ket
V_A\ket{H}_B)$, $\ket{\phi^\pm}=\frac1{\sqrt2}(\ket
H_A\ket{H}_B\pm\ket V_A\ket{V}_B)$ represent an entangled basis for
the four dimensional Hilbert space describing the polarization of two
photons. They can be grouped into the singlet state $\ket{\psi^-}$,
generating the antisymmetric subspace, and the triplet states,
$\{\ket{\psi^+}$, $\ket{\phi^+}$, $\ket{\phi^-}\}$, that generate the
symmetric subspace, where the symmetry is referred to the exchange of
the two photons \cite{matt96prl}. The beam splitter can be used to
discriminate between the symmetric and antisymmetric subspaces.
Indeed if two photons in the singlet state $\ket{\psi^-}$ impinge
simultaneously on a 50/50 BS, they will always emerge on different
outputs of the BS due to quantum interference. Conversely, for any
state orthogonal to $\ket{\psi^-}$ (thus belonging to the symmetric
subspace) the two photons will be found in the same output mode.

The setup adopted in the experiment is shown in Fig.
\ref{fig:setup}. To observe the appearance of the bosonic coalescence
for input symmetric states, we varied the relative delay between the
two photons and hence their corresponding temporal superposition on
the ULWBS \cite{hong87prl}. We first tested the Hong-Ou-Mandel (HOM)
\cite{hong87prl} effect with separable states, by placing two PBS in
the $k_A$ and $k_B$ modes (see Fig. \ref{fig:setup}). We report in
Figure \ref{fig:osc}a) the coincidence counts as a function of the
slit delay $\Delta x$ for two photons in the input state $\ket{HH}$.
The measured visibility, defined as $V_{exp}=|\frac{\mathcal
C_{0}-\mathcal C_{int}}{\mathcal C_{0}}|$, where $\mathcal C_0$ and
$\mathcal C_{int}$ correspond respectively to the coincidence rate
outside interference (i.e. with $\Delta x$ larger than the photon
coherence length) and inside interference ($\Delta
x=0$). The measured visibility is $V=0.937\pm0.009$. We performed the
same measurement with the input states $\ket{VV}$ and $\ket{++}$
obtaining $V=0.926\pm0.012$ and $V=0.954\pm0.011$ respectively. We
also tested the interference with entangled states. When the photons
arrive simultaneously on the ULWBS ($\Delta x=0$ in the figure) we
measured for the triplet (singlet) a dip (peak) in the coincidence
counts as expected (see Figure \ref{fig:osc}b)). The measured
visibility  are $V_{\ket{singlet}}=0.930\pm0.005$ and
$V_{\ket{triplet}}=0.929\pm0.005$. We attribute the slight
discrepancy observed between the theoretical \footnote{Since our BS
has different transmission ($T$) and reflection ($R$) coefficients
for the $H$ and $V$ polarization, the maximum expected visibilities
are $V_\psi = 2\frac{\sqrt{T_HT_VR_HR_V}}{T_HT_V +
R_HR_V}\simeq0.989$ for $|\psi^\pm\rangle$ and $V_\phi
=2\frac{T_HR_H+T_VR_V}{T_H^2 + R_H^2 + T_V^2 + R_V^2}\simeq0.974$ for
$|\phi^\pm\rangle$ } and experimental values to a partial spectral
distinguishability between the photons on modes $k_A$ and $k_B$: this
could be reduced by using narrower bandwidth detection filters.

Let's now analyze the behavior of the different entangled states. The
temporal delay was set at $\Delta x=0$ and the source was tuned to
generate the entangled state $\ket{\psi^+}$. By inserting on mode
$k_B$ a half waveplate (HWP) with the optical axis oriented at an
angle $\theta$ with respect to the vertical direction,
 the following states are generated: $ -\cos2\theta\ket{\psi^-}+\sin2\theta\ket{\phi^+}$.
In this case, the expected coincidence rates between detectors $D_C$
and $D_D$ after the beam splitter is $\mathcal N_0[1+\widetilde
V\cos4\theta]$ where the expected visibility with the given $R_H$ and
$R_V$ can be found to be $\widetilde V=0.973$. The experimental results are
shown in Fig. \ref{fig:osc}c), yielding a visibility $V
= 0.962 \pm 0.018$.

When the source is tuned to generate the entangled state
$\ket{\psi^i}=\frac1{\sqrt2}(\ket H_A\ket{V}_B-i\ket V_A\ket{H}_B)$,
by using a quarter waveplate (QWP) rotated at $\theta'$ on mode
$k_A$, the state is found to be
$\cos^2\theta'\ket{\psi^-}-i\sin^2\theta'\ket{\psi^+}+\frac{e^{-i\frac{\pi}4}}{\sqrt2}\sin2\theta'\ket{\phi^i}$
with $\ket{\phi^i}=\frac{1}{\sqrt2}(\ket{HH}+i\ket{VV})$. For the
sake of simplicity, by assuming polarization independent
reflectivity, the coincidence rates expected with a beam splitter
with reflectivity $R$ is $\mathcal N_0[
1-V_{teo}+2V_{teo}\cos^4\theta']$ where
$V_{teo}=2(1-R)R/(2R^2-2R+1)$. By taking $R=(R_H+R_V)/2$ we will
expect $V_{teo}=0.987$. The theoretical behavior was verified in the
experiment. The corresponding fringe pattern, with visibility
$V=0.951\pm0.008$, is shown in Fig. \ref{fig:osc}d). These results
demonstrate the high overlap between the interfering modes $k_A$ and
$k_B$ and show that ULWBS may be used as an appropriate tool for the
manipulation of polarization encoded qubit.
\begin{figure}
\centering
\includegraphics[width=6.5cm]{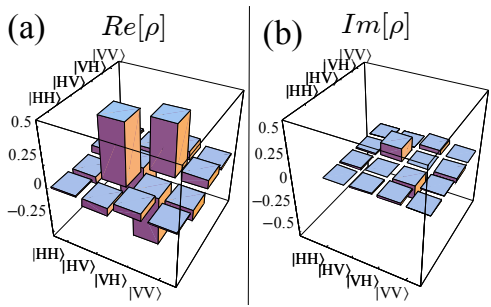}
\caption{Quantum state tomography of a filtered singlet state on the two output mode $k_C$ and $k_D$.
Real (a) and imaginary (b) part of the experimental density matrix of the filtered state.}
\label{fig:tomo}
\end{figure}

As a final experimental characterization we adopted the
ULWBS to carry out the projection on the singlet
subspace. We injected into the
ULWBS the separable state $\ket{H}_A\otimes\ket{V}_B$
and analyzed the output state when two photons emerge on the two
modes $k_C$ and $k_D$: the expected state reads $\ket{\psi^-}_{CD}$.
We performed the quantum state tomography  \cite{jame01pra} of the
output state conditioned to the detection of the two photons in
different outputs.
In this case two standard polarization analysis setup were adopted after the ULWBS
(see dashed box in Fig. \ref{fig:setup}).
The experimental density matrix, $\rho_{CD}$,
shown in Figure \ref{fig:tomo} exhibits a low entropy
($S_L=0.071\pm0.018$), a high concurrence ($C=0.941\pm0.015$) and a
high fidelity with the singlet state ($F=0.929\pm0.007$). We observe
that the present scheme achieves {\it a-posteriori} singlet component
filtration, i.e., conditioned to the detection of one photon per
output mode. Recently, a heralded entanglement filter, based on two
auxiliary photons and an interferometric scheme, has
been reported by adopting a bulk optical scheme in Ref.
\cite{okam09sci}.

In summary, we reported on the realization and quantum
optical characterization of a
femtosecond-laser-written directional coupler, acting as an
integrated beam splitter. The experimental results demonstrate the
suitability of this method to manipulate qubit encoded in the
polarization of photon states. In order to achieve a complete
handling of the polarization degree of freedom the next step will consist
in the realization of integrated tunable waveplates and polarizing beam splitters.
By combining these tools with integrated sources of photon pairs
and, possibly, with integrated detectors,
the realization of pocket quantum optics lab, available for
optical quantum sensing, computing and metrology may become a reality in the near future.

\acknowledgements
 This work was supported by FARI project and Finanziamento Ateneo 2009 of Sapienza Universit{\aac} di
 Roma and Finanziamento Ateneo 2009 of Politecnico di Milano.


\begin{thebibliography}{10}
\providecommand{\url}[1]{\texttt{#1}}
\providecommand{\urlprefix}{URL }
\providecommand{\eprint}[2][]{\url{#2}}

\bibitem{gisi02rmp}
N.~Gisin, \emph{et~al.}, Rev. Mod. Phys. \textbf{74}, 145 (2002).

\bibitem{duan01nat}
L.-M. Duan, \emph{et~al.}, Nature \textbf{414}, 413 (2001).

\bibitem{ladd10nat}
T.~D. Ladd, \emph{et~al.}, Nature \textbf{464}, 45 (2010).

\bibitem{kok07rmp}
P.~Kok, \emph{et~al.}, Rev. Mod. Phys. \textbf{79}, 135 (2007).

\bibitem{dowl08cph}
J.~Dowling, Contemporary Physics \textbf{49}, 125 (2008).

\bibitem{poli08sci}
A.~Politi, \emph{et~al.}, Science \textbf{320}, 646 (2008).

\bibitem{poli09sci}
A.~Politi, J.~C.~F. Matthews, and J.~L. O'Brien, Science \textbf{325}, 1221
  (2009).

\bibitem{matt09npho}
J.~Matthews, \emph{et~al.}, Nat. Photonics \textbf{3}, 346 (2009).

\bibitem{smit09ope}
B.~J. Smith, \emph{et~al.}, Optics Express \textbf{17}, 13516 (2009).

\bibitem{kris10npho}
R.~Krischek, \emph{et~al.}, Nat. Photonics \textbf{4}, 170 (2010).

\bibitem{walt05nat}
P.~Walther, \emph{et~al.}, Nature (London) \textbf{434}, 169 (2005).

\bibitem{kawachi90oqe}
M.~Kawachi, Opt. Quantum Electron. \textbf{22}, 391 (1990)

\bibitem{johlen00jlt}
D.~Johlen, \emph{et al.}, J. Lightwave Technol. \textbf{18}, 185 (2000).

\bibitem{inoue97jlt}
Y. Inoue, \emph{et al.}, J. Lightwave Technol. \textbf{15}, 1947 (1997).


\bibitem{gattass08np}
R.R.Gattass and E.~Mazur, Nat. Photonics \textbf{2}, 219 (2008).

\bibitem{dellavalle09joa}
G.~{Della Valle}, R.~Osellame, and P.~Laporta, Journal of Optics A \textbf{11},
  013001 (2009).

\bibitem{osellame06jstqe}
R.~Osellame, \emph{et~al.}, Selected Topics in Quantum Electronics, IEEE
  Journal of \textbf{12}, 277  (2006), ISSN 1077-260X.

\bibitem{osel03jos}
R.~Osellame, \emph{et~al.}, Journal of Optical Society of America B
  \textbf{20}, 1559 (2003).

\bibitem{mars09ope}
G.~D. Marshall, \emph{et~al.}, Optics Express \textbf{17}, 12546 (2009).

\bibitem{cheng09ope}
G.~Cheng, \emph{et~al.}, Opt. Express \textbf{17}, 9515--9525 (2009).

\bibitem{shimotsuma03prl}
Y.~Shimotsuma, \emph{et~al.}, Phys. Rev. Lett. \textbf{24}, 247405 (2003).

\bibitem{ams08jstqe}
M.~Ams, \emph{et~al.}, Selected Topics in Quantum Electronics, IEEE Journal of
  \textbf{14}, 1370  (2008).

\bibitem{eaton08ope}
S.~M. Eaton, \emph{et~al.}, Opt. Express \textbf{16}, 9443 (2008).

\bibitem{si1} See    supplementary    material  for   details.

\bibitem{matt96prl}
K.~Mattle, \emph{et~al.}, Phys. Rev. Lett. \textbf{76}, 4656 (1996).

\bibitem{hong87prl}
C.~K. Hong, Z.~Y. Ou, and L.~Mandel, Phys. Rev. Lett. \textbf{59}, 2044 (1987).

\bibitem{jame01pra}
D.~F.~V. James, \emph{et~al.}, Phys. Rev. A \textbf{64}, 052312 (2001).

\bibitem{okam09sci}
R.~Okamoto, \emph{et~al.}, Science \textbf{323}, 483  (2009).

\end{thebibliography}
\end{document}


\title{Supplementary informations: Polarization entangled state measurement on a chip}
\author{Linda Sansoni}
\affiliation{\fisicarm}
\author{Fabio Sciarrino}
\affiliation{\fisicarm}
\affiliation{\ino}
\author{Giuseppe Vallone}
\affiliation{\cfermi}
\affiliation{\fisicarm}
\author{Paolo Mataloni}
\affiliation{\fisicarm}
\affiliation{\ino}
\author{Andrea Crespi}
\affiliation{\ifn}
\affiliation{\fisicami}
\author{Roberta Ramponi}
\affiliation{\ifn}
\affiliation{\fisicami}
\author{Roberto Osellame}
\affiliation{\ifn}
\affiliation{\fisicami}

\maketitle

\section{Device fabrication}

In our waveguide writing setup we employed a Yb:KYW cavity-dumped
mode-locked oscillator \cite{osellame06jstqe}, delivering $300 $fs,
$1 \mu$J pulses at $1030 $nm wavelength, with $1 $MHz repetition
rate.
Laser pulses of $240 $nJ were focused into the EAGLE2000 (Corning Inc.) substrate using
a 0.6 NA microscope objective. Sample translation at a constant
writing speed of $40$mm/s was enabled by high precision three-axes
air bearing stages (Aerotech Fiber-Glide 3D).

ULWBS devices were fabricated directly buried inside the glass substrate at a constant depth of $170 \mu$m.

Permanent fiber coupling of the ULWBS was obtained by gluing two
pairs of fibers at the input and at the output ports and this introduced additional coupling losses lower that 0.6 dB/facet. Each pair of
fibers was first inserted into a quartz double-ferrule (Friedrich \&
Dimmock, inc. - NJ U.S.A.), which allows to maintain a fixed distance
of $250 \mu m$ between the two fibers.

\section{Waveguide birefringence characterization}

\begin{figure}[t]
	\centering
	\includegraphics[width=8.5cm]{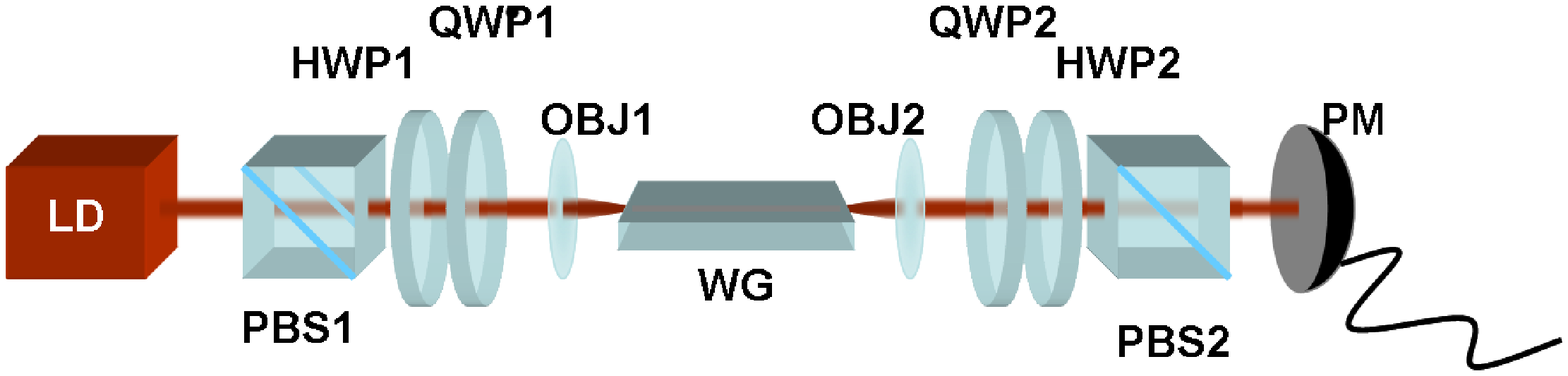}
	\caption{Setup for birefringence measurement. LD: laser diode emitting at $806 $nm wavelength and with bandwidth $\Delta\lambda=2$ nm. PBS1 and PBS2: polarizing beam splitters. HWP1 and HWP2: half-wave plates. QWP1 and QWP2: quarter-wave plates. OBJ1 and OBJ2: objectives for injecting light into the waveguide (WG) and collecting transmitted light at the output. PM: power meter (PM).}
	\label{fig:birefr}
\end{figure}

The birefringence of the fabricated waveguides was characterized by the setup shown in Fig. \ref{fig:birefr}.
Six different polarization eigenstates ($\ket{H}, \ket{V}, \ket{+}, \ket{-}, \ket{L}, \ket{R}$)  were launched into the waveguide. The eigenstates were selected by rotating HWP1 and, if necessary, by adding also QWP1. For each of them the projections of the output state on all the same six eigenstates were measured (acting with waveplates HWP2 and QWP2), thus allowing the calculation of the normalized Stokes vectors for the output state.\\
It was assumed as hypothesis that the waveguide is analogous to a uniaxial birefringent material with the optical axis tilted by an angle $\theta$, which introduces a dephasing $\delta$ between the two polarizations corresponding to the ordinary and extraordinary refractive index. As a consequence, a normalized Mueller matrix of the form:
$$
\footnotesize \left[\begin{array}{cccc}
1 & 0 & 0 & 0 \\
0 & cos^2 2\theta + sin^2 2\theta cos \delta & sin 2\theta cos 2 \theta \left( 1 - cos \delta \right) & - sin 2 \theta sin \delta \\
0 & sin 2\theta cos 2 \theta \left( 1 - cos \delta \right) & sin^2 2\theta + cos^2 2\theta cos \delta & cos 2 \theta sin \delta \\
0 & sin 2 \theta sin \delta & - cos 2 \theta sin \delta & cos \delta
 \end{array}\right]
$$
was fitted to the experimental data in order to determine the transfer function from the input to the output Stokes vectors. $\delta$ and $\theta$ were the free parameters in the fitting process.

However, a single measurement of $\delta$ cannot be used for direct determination of birefringence, since ambiguity occurs with multiple-order birefringence. Thus, the sample was cut to different lengths and  $\delta$ and $\theta$ were measured as described above for each length. This allowed to remove the ambiguity and yielded the actual waveguide birefringence.

In the ULW waveguides used for the experiments we found a fast axis aligned with the TE polarization and a birefringence value $B = n_{eff,TM} - n_{eff,TE} = 7 \times 10^{-5}$.

\section{Source}
Polarization entangled photon pairs were generated via spontaneous parametric down
conversion in a $1.5$mm $\beta$-barium borate (BBO)
crystal cut for type-II non-collinear phase matching
\cite{kwia95prl}, pumped by a CW laser diode with power $P = 50$mW and low coherence time ($\tau_p<1$ps).
Degenerate photons at wavelength $\lambda = 806$nm were detected
within a spectral bandwidth $\Delta\lambda =6$nm, as determined by
the interference filters.
Waveplates and polarizing beam splitter (PBS) were optionally inserted in path $k_A$ and $k_B$ to prepare
different input states. The setup shown in the dashed box have been only inserted in the quantum state
tomography of the filtered state. Suitable BBO crystal compensators (C)
and polarization controllers (PC) were respectively used on
each photon path to compensate temporal walk-off and ensure polarization
maintenance of the photons on the ULWBS. We could measure a coincidence rate
$C_{source} = 3$kHz when photon radiation was
coupled into two single mode fibers and
directly connected to detectors.
A delay line $\Delta x$ in the $k_A$ arm
enabled temporal delay variation between the two input photons.
\section{Data analysis}
The data shown in Figure 4 were obtained by adding the measurement device shown in the dashed box placed on the right of Figure 2, as explained in the manuscript.
The additional optical component (HWP, QWP and PBS) and a more critical polarization compensation are mainly responsible of the measured fidelity of the reconstructed state. Let's try to evaluate all the imperfection contribution.
We may attribute the 7\% missing fidelity to the following: $\sim3\%$ are due to the imperfection in the source and integrated
chip (a fidelity of 97\% would be compatible with the average visibility reported in Figure 3), $\sim2\%$ of missing fidelity
is due to a non perfect polarization compensation (in fact, the phase between the states $\ket{HV}$ and $\ket{VH}$ of the
reconstructed state has a residual imaginary component as shown in Figure 4) and $\sim 2\%$ may be attributed to the tomographic measurement setup.